\documentclass[
10pt,
twocolumn,
twoside,
superscriptaddress,
nofootinbib,
amsmath,
amssymb,
aps,
prl,
]{revtex4-2}

\usepackage{graphicx}
\usepackage{microtype}
\usepackage{dcolumn}
\usepackage{bm}
\usepackage{xcolor}

\makeatletter
\def\tagform@#1{(\textcolor{magenta}{#1})}
\makeatother
\usepackage[colorlinks=true,linkcolor=magenta,citecolor=magenta]{hyperref}
\usepackage{nicefrac}

\makeatletter\renewcommand\paragraph{\@startsection{paragraph}{4}{\parindent}{0.6ex plus 0.2ex}{-0.01pt}{\normalfont\bfseries\itshape}}\makeatother
\begin{document}

\title{\color{teal}Conformal Minimal Models from Functional Dimensional Regularization}

\author{A. Codello}
\email{alessandro.codello@unive.it}
\affiliation{DSMN, Ca' Foscari University of Venice, Via Torino 155, 30172 - Venice, Italy}
\affiliation{IFFI, Facultad de Ingenier\'ia, Universidad de la Rep\'ublica, J.H.y Reissig 565, 11000 Montevideo, Uruguay}

\begin{abstract}
Can the Renormalization Group (RG), with no input from conformal symmetry, reconstruct the data of the two-dimensional
Conformal Minimal Models (CMM)? We show that it can, with a one-loop functional flow that fits in two lines. Functional Dimensional Regularization
(FDR), in the essential scheme at second order of the derivative expansion, reduces to one equation for the potential and one
constraint for the field redefinition that keeps the wave-function renormalization fixed. Removing the redundant operators turns the
anomalous dimension into an eigenvalue and makes the space of scaling solutions one-dimensional, as in the local potential
approximation. Scaling solutions are then spikes of a one-parameter shooting problem in every $2\le d\le4$: stacked in $d$, they form
the {\color{teal}{\it Cordillera of} CMM}, whose ridges branch off the Gaussian at each upper critical dimension $d_c=2n/(n-1)$ and end, in $d=2$, on the minimal model
$\mathcal{M}_{n+1,n+2}$. The spin-zero spectrum stays real and evolves continuously from the Gaussian operators $\varphi^k$ to the
Kac primaries. Directly in $d=2$, $\eta$ and $\nu$ are within $8\%$ and $6.4\%$ of the exact values for {\tt Ising}, and the errors
decrease along the series, down to $0.7\%$ and $0.02\%$ for $\mathcal{M}_{11,12}$; above $d=2$ the spectrum agrees with Conformal
Bootstrap data in fractional dimensions. To our knowledge, this is the first RG description of all the CMM, from their upper critical
dimensions down to two dimensions.
\end{abstract}

\maketitle

\paragraph*{\color{teal}\mbox{Introduction.---}}
\begin{figure*}[t]
    \centering
    \includegraphics[width=\textwidth]{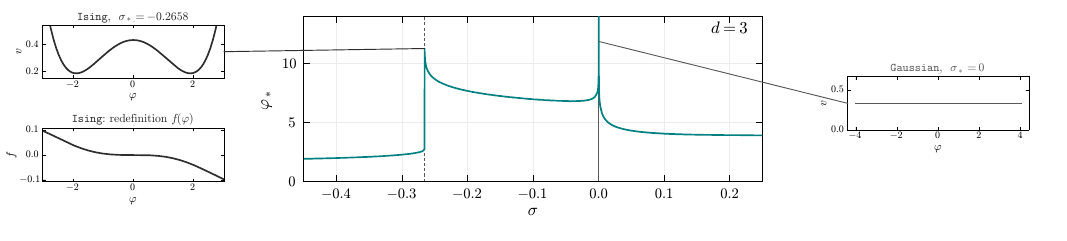}\\[-1mm]
    \includegraphics[width=\textwidth]{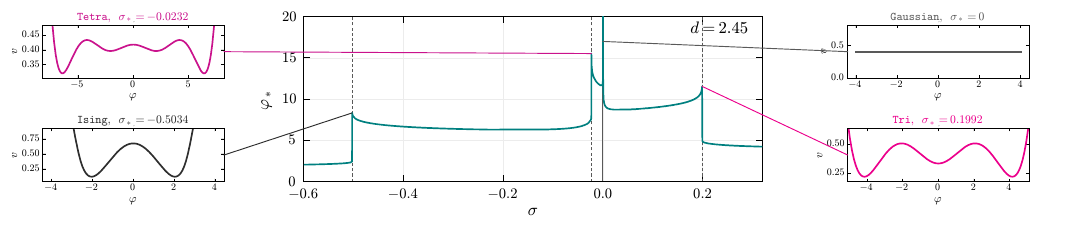}\\[-1mm]
    \includegraphics[width=\textwidth,trim=0 10 0 0,clip]{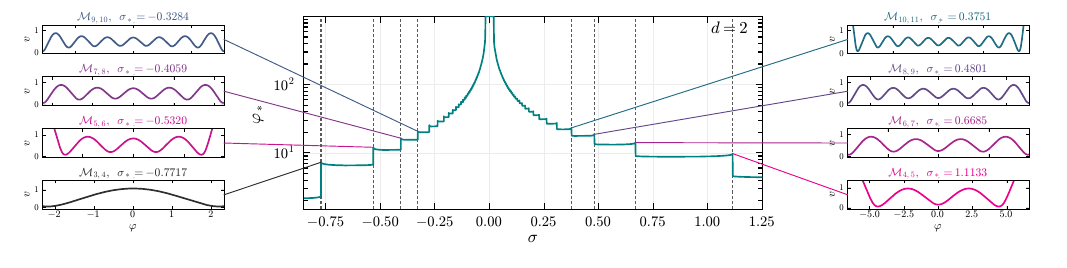}
    \caption{Spike plots of essential FDR-DE2 in $d=3$, $2.45$ and $2$ (top to bottom; log scale in $d=2$): the field $\varphi_*$ at
    which the solution shot from $\varphi=0$ with $v''(0)=\sigma$ breaks down. Dashed: fixed points; solid: Gaussian. Side panels: the
    fixed-point potentials $v(\varphi)$ obtained by shooting at $\sigma_*$, joined to their spike (top: also the field redefinition
    $f(\varphi)$ of {\tt Ising}). In $d=2.45$ {\tt Tetra} ($\sigma_*=-0.023$) has separated from the Gaussian spike, while {\tt Penta},
    born at $d_c=5/2$, is at $\sigma_*=4.6\times10^{-4}$, unresolved. In $d=2$ all multicritical points form two staircases on the two
    sides of $\sigma=0$, labelled by their minimal model; the minima are evenly spaced, $\Delta\varphi\simeq4.44$. The $d=2$ plot is
    cut at $\varphi_*=10^3$, the integration range.}
    \label{EssentialSpike}
\end{figure*}

The complete solution of the Conformal Minimal Models (CMM) $\mathcal{M}_{m,m+1}$, $m=3,4,5,\dots$
\cite{Belavin:1984vu,Friedan:1983xq,DiFrancesco:1997nk} is a crowning achievement of two-dimensional conformal field theory (CFT).
They describe the multicritical points of a single scalar field in $d=2$: the Landau--Ginzburg theory with a $\varphi^{2n}$ potential
flows to $\mathcal{M}_{n+1,n+2}$ \cite{Zamolodchikov:1986db}, so that {\tt Ising} ($n=2$) is $\mathcal{M}_{3,4}$, {\tt Tricritical}
is $\mathcal{M}_{4,5}$, and so on. All their critical exponents are known exactly; for example the anomalous dimension of the leading
$\mathbb{Z}_2$-odd operator is $\eta=\frac{3}{m(m+1)}$.
These exact results rest on conformal symmetry. The Renormalization Group (RG), on the other hand, knows nothing about it, and yet
the CMM are fixed points of the RG flow of a scalar field. Can the RG alone reconstruct them? In practice two dimensions are the
epitome of non-perturbativity: the $\varepsilon$-expansion \cite{Wilson:1971dc,Wilson:1973jj} does not reach $d=2$. The Conformal
Bootstrap (CB) \cite{Poland:2018epd} has followed the {\tt Ising} CFT in fractional dimensions \cite{El-Showk:2013nia} all the way
down to $d=2$, the violation of unitarity in non-integer $d$ turning out to have a negligible effect \cite{Cappelli:2018vir}; more
recently, it has tracked the whole spin-zero spectrum, level repulsions included, from $d=4$ down to $d\simeq2.6$ \cite{Henriksson:2022gpa} (see also Ref.~\cite{Bonanno:2022ljl} for $3\le d<4$), and has
located the {\tt Tricritical} CFT below $d=3$ \cite{Henriksson:2025tri}. These spectra in fractional dimensions are the benchmark for
any RG description, and they provide the key test of this Letter. Functional RG studies \cite{Wetterich:1989xg,Morris:1993qb,Dupuis:2020fhh}
followed the scaling solutions in continuous dimension \cite{Ballhausen:2003gx,Codello:2012sc} and, at second order of the derivative
expansion (DE2) \cite{Morris:1994jc,Defenu:2017dec}, found the multicritical fixed points in $d=2$, first qualitatively and then
quantitatively. What has been missing is a single RG equation that contains the whole tower of multicritical classes, follows each of
them from its upper critical dimension down to $d=2$, and keeps track of the full spectrum along the way.

The main obstacle is the presence of redundant operators \cite{Wegner:1974}, those that can be removed by a field redefinition. Their
eigenvalues carry no physical information, but in truncated flows they mix with the physical ones: levels of the same symmetry
collide and become complex, and \emph{spectrum continuity}, the smooth evolution of each scaling dimension from $d_c$ to $d=2$, is lost.
This happens in the non-perturbative RG at DE2 and DE4 \cite{Vacca:2026prep} and, we expect, in any functional RG that keeps redundant operators.

\begin{figure*}[t]
    \centering
    \includegraphics[width=\textwidth,trim=0 19 0 0,clip]{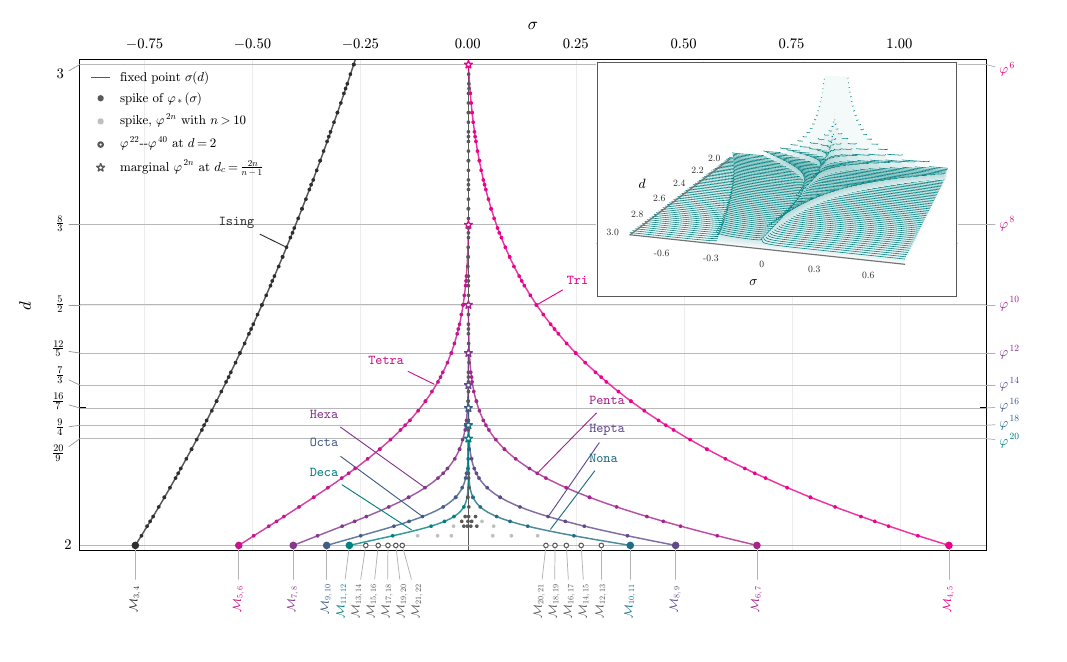}
    \caption{The {\color{teal}{\it Cordillera of} CMM}. Inset: spike plots $\varphi_*(\sigma)$ stacked in $d$ from $3$
    (front) to $2$ (back), each slice broken at its spikes. Main panel: its shadow on the $(\sigma,d)$ plane, the positions of the spikes of
    the slices, coloured by the class whose line of fixed points $\sigma_*(d)$ (lines) they sit on; grey dots: classes with $n>10$, open
    circles: $\varphi^{22}$--$\varphi^{40}$ at $d=2$. Each ridge leaves the Gaussian line $\sigma=0$ at $d_c=\frac{2n}{n-1}$ (horizontal
    lines, values on the left), where $\varphi^{2n}$ is marginal (stars, labelled on the right), and ends at $d=2$ on its minimal model
    $\mathcal{M}_{n+1,n+2}$ (bottom).}
    \label{CordilleraShadow}
\end{figure*}

\begin{figure*}[t]
    \centering
    \includegraphics[width=\textwidth,trim=0 6 0 0,clip]{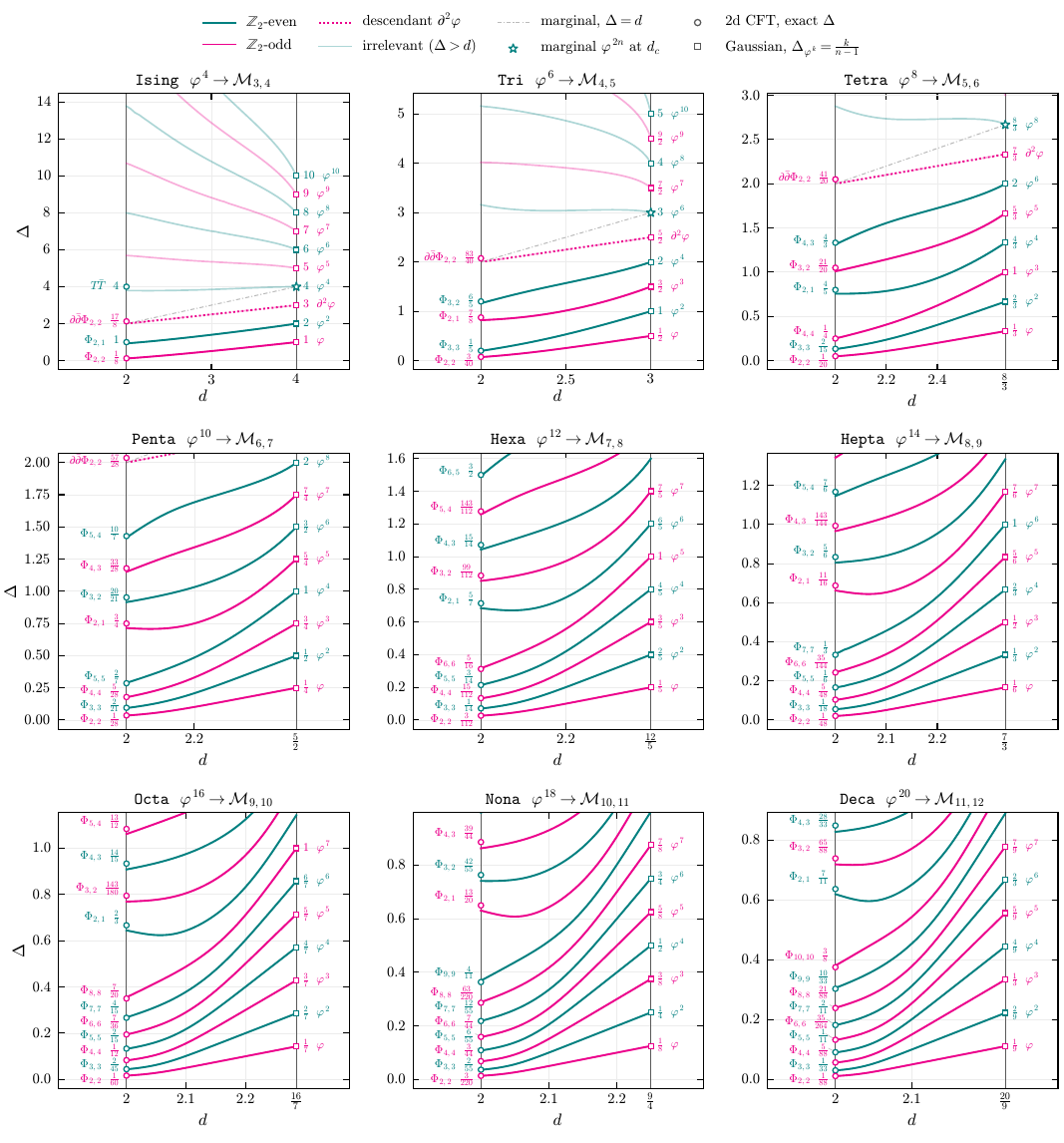}
    \caption{Spectrum continuity: scaling dimensions $\Delta=d+\theta$ of the spin-zero $\mathbb{Z}_2$-even (teal) and
    $\mathbb{Z}_2$-odd (magenta) operators of the first nine classes from $d=2$ to $d_c$, in the range of the first four even and four
    odd operators. Left: the minimal-model operator each level flows to, with its exact $\Delta$ (Kac primaries $\Phi_{r,s}$,
    $\Delta=2h_{r,s}$; the descendant $\partial\bar\partial\Phi_{2,2}$, $\Delta=\Delta_\varphi+2$; $T\bar T$ for the {\tt Ising}
    $\varphi^4$). Right: the Gaussian operator $\varphi^k$ with $\Delta=\frac{k}{n-1}$; the exact $d_c$ is marked on the $d$ axis.
    All classes are computed on the same uniform grid of 40 points in $x=(d-2)/(d_c-2)$, refined towards $d_c$; every curve ends at
    its Gaussian value. Lines are light where the operator is irrelevant ($\Delta>d$); dotted: the descendant $\partial^2\varphi$;
    dash-dotted: the marginal line $\Delta=d$; star: the marginal $\varphi^{2n}$ at $d_c$ (above the range shown for $n\ge5$).}
    \label{CMMspectral}
\end{figure*}

In this Letter we show that a remarkably simple RG equation overcomes these difficulties at once. It combines two ingredients: Functional Dimensional Regularization (FDR) \cite{FDRLetter,FDR}, which is a regulator-free one-loop functional RG, and the essential RG
\cite{Baldazzi:2021ydj}, in which the redundant operators are removed from the start by a running field redefinition. At DE2 the
resulting flow consists of just two equations, Eqs.~\eqref{betav}--\eqref{betaf} below, each shorter than one line, which know nothing
about conformal symmetry. Yet they produce: all the multicritical fixed points $\varphi^{2n}$ in every dimension $2\le d\le d_c$,
organized in the {\color{teal}{\it Cordillera of} CMM}, whose ridges join each upper critical dimension to its minimal model (Fig.~\ref{CordilleraShadow}); a real
spin-zero spectrum that interpolates continuously between the Gaussian operators and the Kac primaries (Fig.~\ref{CMMspectral}) and
agrees with the bootstrap spectra in fractional dimensions \cite{Cappelli:2018vir,Henriksson:2022gpa,Henriksson:2025tri}; and
critical exponents in $d=2$ whose errors, below $8\%$ for {\tt Ising}, decrease along the series to well below one per cent
(Table~\ref{tab:d2}). To our knowledge, this is the first time that an RG equation reproduces, qualitatively and quantitatively, the
whole family of CMM from their upper critical dimensions down to two dimensions.

The Letter is organized as follows. We first introduce FDR and its essential version, and test the resulting equations on the
Wilson--Fisher fixed point in $d=3$. We then solve them at the functional level in every dimension, building the
{\color{teal}{\it Cordillera of} CMM}, and finally compute the spin-zero spectrum and compare it with the exact two-dimensional
results, the CB and the $\varepsilon$-expansion. Technical details are collected in the End Matter.

\paragraph*{\color{teal}\mbox{Essential FDR.---}}
\begin{figure}[t]
    \centering
    \includegraphics[width=\columnwidth,trim=0 8 0 0,clip]{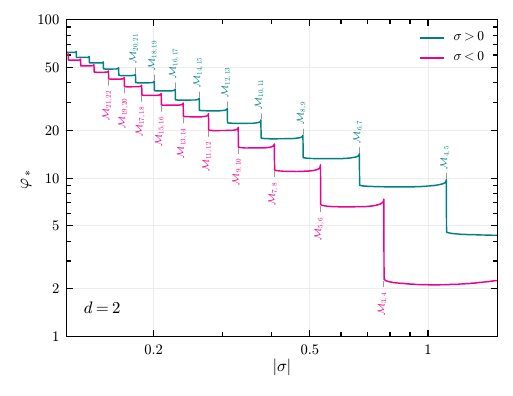}
    \caption{The CMM stairway: both branches of the $d=2$ spike plot against $|\sigma|$ on log-log axes ($\sigma>0$ teal,
    $\sigma<0$ magenta). Every step ends on the spike of a minimal model, from $\mathcal{M}_{3,4}$ to $\mathcal{M}_{21,22}$
    ($\sigma>0$ above, $\sigma<0$ below); the two staircases interleave and continue towards the Gaussian point.}
    \label{EssentialStairway}
\end{figure}

FDR builds on dimensional regularization \cite{Bollini:1972ui,tHooft:1973mfk} and on its relation to the functional RG
\cite{Baldazzi:2020vxk}. Dimensional regularization sees only the poles of the critical dimension in which one works; FDR subtracts
the poles of \emph{all} critical dimensions at once. Expanding the effective action on a basis of operators,
$\Gamma=\sum_i g_i\,\mathcal{O}^i$, the FDR beta functions are
\begin{equation}\label{master}
\beta_i(d)=\sum_{d_c}\mu^{d-d_c}\,\beta_i^{\rm DR}(d_c)\,,
\end{equation}
where $\beta_i^{\rm DR}(d_c)$ is the dimensional-regularization beta function of $g_i$ in the critical dimension $d_c$, so that the
full one-loop flow is $\partial_t\Gamma=\beta_\Gamma$ with $\beta_\Gamma=\sum_i\beta_i\,\mathcal{O}^i$. The flow is regulator free,
and the dimension $d$ enters only through the scaling terms: this is what makes the limit $d\to2$ accessible.

In the standard FDR-DE2 scheme \cite{FDR} the scaling solutions depend on two parameters, the mass $\sigma=v''(0)$ and the
anomalous dimension $\eta$, and appear as isolated peaks in a landscape over the $(\sigma,\eta)$ plane \cite{FDR2d}. In $d=2$ this landscape
contains {\tt Ising}, {\tt Tricritical} and {\tt Tetracritical}, with $\eta$ within $3$--$7\%$ of the exact values. However,
the spin-zero spectrum contains redundant eigenvalues, which collide with the essential ones: $\omega$ becomes complex near $d=2$
(End Matter, Fig.~\ref{SpectrumDE2}).

The way out is the essential RG \cite{Baldazzi:2021ydj}: the field itself is allowed to run, $\partial_t\phi=\Phi[\phi]$, and the
flow equation becomes
\begin{equation}\label{essential}
\partial_t\Gamma+\int\frac{\delta\Gamma}{\delta\phi}\,\Phi=\beta_\Gamma\,.
\end{equation}
Couplings that can be changed by a field redefinition are inessential: they carry no physical information, and the freedom in $\Phi$
can be used to fix them. At DE2, with $\Phi=F(\phi)$, one can keep the wave-function renormalization equal to one, so that
$\Gamma=\int\{\frac12(\partial\phi)^2+V(\phi)\}$ along the flow. Then
$\int\frac{\delta\Gamma}{\delta\phi}\Phi=\int\{\frac12(\partial\phi)^2\,2F'+V'F\}$, and Eq.~\eqref{essential} splits into the
essential flow $\partial_tV+V'F=\beta_V|_{Z=1}$ and the constraint $2F'=\beta_Z|_{Z=1}$. With FDR, in dimensionless variables, the
flow of the potential is
\begin{equation}\label{betav}
\colorbox{lightgray!22}{$\displaystyle\beta_{v}=-dv+\big(\tfrac{d-2}{2}\varphi-f \big)v'+e^{-v''}$}\,,
\end{equation}
and the field redefinition is fixed by the constraint
\begin{equation}
\colorbox{lightgray!22}{$\displaystyle 0=f'+\tfrac{1}{12}(v''')^{2}\,e^{-v''}$}\,.  \label{betaf}
\end{equation}
Equations~\eqref{betav}--\eqref{betaf} are all there is: they contain no regulator function and no free parameter, and they know
nothing about conformal symmetry. The functionals $\beta_V$ and $\beta_Z$ were derived in Ref.~\cite{FDR}; as a check, these
equations can also be obtained by {\it distillation} from the analogous DE2 essential RG equations of the non-perturbative RG
\cite{Baldazzi:2021ydj}. We have rescaled away the factors of $4\pi$ ($\varphi\to\varphi/\sqrt{4\pi}$, $\sigma\to4\pi\sigma$).
The standard scheme corresponds to the linear choice $f=-\frac{\eta}{2}\varphi$, which gives back the familiar scaling term
$\frac{d-2+\eta}{2}\varphi v'$. In the essential scheme $f(\varphi)$ is instead determined by Eq.~\eqref{betaf}: $\eta$ is no longer
an input, and it emerges as an eigenvalue of the linearized flow. The functional theory space is therefore one-dimensional,
parametrized by $\sigma\equiv v''(0)$ as in the local potential approximation (LPA), yet it contains the anomalous dimension. This is
the key simplification.

To make contact with the usual perturbative description, we first expand in the field,
$v=\sum_n\frac{\lambda_n}{n!}\varphi^n$ and $f=\sum_n\frac{\gamma_n}{n!}\varphi^n$. Equation~\eqref{betav} becomes an infinite
system of beta functions $\beta_n=\partial_\varphi^n\beta_v|_{\varphi=0}$, the first of which are
\begin{eqnarray}
\beta_{1} & = &-\gamma _1 \lambda _1-\tfrac{d}{2}\lambda _1-\lambda _1-e^{-\lambda _2} \lambda _3\nonumber\\
\beta_{2} & = & -\gamma _2 \lambda _1-2 \gamma _1 \lambda _2-2 \lambda _2+e^{-\lambda _2} (\lambda _3^2-\lambda _4)\nonumber\\
\beta_{3} & = & -\gamma _3 \lambda _1-3 \gamma _1 \lambda _3-3 \gamma _2 \lambda _2+\tfrac{d}{2}\lambda _3-3 \lambda _3\nonumber\\
&&-e^{-\lambda _2} (\lambda _3^3-3\lambda _4 \lambda _3)\nonumber\\
               & \vdots &\label{fexpbeta}
\end{eqnarray}
and the constraint~\eqref{betaf} an infinite set of conditions
\begin{eqnarray}
0 & = & \gamma _1+\tfrac{1}{12}  \lambda _3^2 \,e^{-\lambda _2}\nonumber\\
0 & = & \gamma _2-\tfrac{1}{12}  \big(\lambda _3^3 -2 \lambda _4 \lambda _3 \big)e^{-\lambda _2}\nonumber\\
0 & = & \gamma _3+\tfrac{1}{12}  \big(\lambda _3^4+2 \lambda _5 \lambda _3+2 \lambda _4^2-5 \lambda _4 \lambda _3^2\big) e^{-\lambda _2}\nonumber\\
   & \vdots &\label{fexpcond}
\end{eqnarray}
The conditions~\eqref{fexpcond} are linear in the $\gamma$'s and are solved order by order, $\gamma_n=\gamma_n(\lambda)$. Inserting them
into Eq.~\eqref{fexpbeta} gives a closed system of beta functions for the essential couplings $\lambda_n$ alone: the field redefinition has been
traded for a modification of the flow of the potential. Since $\beta_n$ involves the couplings up to $\lambda_{n+2}$, we truncate at
order $\varphi^{N_{\rm tr}}$, setting $\lambda_n=0$ for $n>N_{\rm tr}$, solve $\beta_1=\dots=\beta_{N_{\rm tr}}=0$ and diagonalize the $N_{\rm tr}\times N_{\rm tr}$ stability matrix
$\partial\beta_i/\partial\lambda_j|_*$. Throughout we write its eigenvalues as $\theta_n=\Delta_n-d$, so that relevant directions have
$\theta_n<0$.

As a first test we apply this procedure in $d=3$ to the Wilson--Fisher fixed point, the representative of the {\tt Ising}
universality class, and increase $N_{\rm tr}$ until the results converge. At order $\varphi^{16}$ the mass $\sigma$ and the exponents $\nu$
and $\eta$ are already stable to four significant digits, and by order $\varphi^{28}$ we find $\sigma_{\tt Ising}=-0.26580$
($-3.3405$ in units with the $4\pi$'s), $\eta = 0.034159$, $\nu = 0.62436$ and $\omega = 0.85027$, with $\omega$ the slowest to
converge. These are very close to the values of the standard
FDR-DE2 equations of Ref.~\cite{FDR}, $\eta = 0.03426$, $\nu = 0.62438$ and $\omega = 0.8554$: the two schemes represent the same truncation, and where the standard scheme is reliable the essential one reproduces it. The first
scaling dimensions agree with the CB values compiled in Ref.~\cite{Henriksson:2025hwi} to within $1\%$ for $\Delta_\sigma=0.5171$
($-0.2\%$) and $\Delta_\epsilon=1.3984$ ($-1.0\%$), and to within $2\%$ for $\Delta_{\epsilon'}=3.850$, $\Delta_{\sigma'}=5.347$ and
$\Delta_{\epsilon''}=7.007$. Near each $d_c$ the equations reproduce the $\varepsilon$-expansion at $O(\varepsilon)$, and for $d_c=4$
also the two-loop $\eta=\frac{\varepsilon^2}{54}$ (End Matter).

\paragraph*{\color{teal}\mbox{The {\it Cordillera of} CMM.---}}
\begin{figure*}[t]
    \centering
    \includegraphics[width=0.49\textwidth,trim=0 16 0 0,clip]{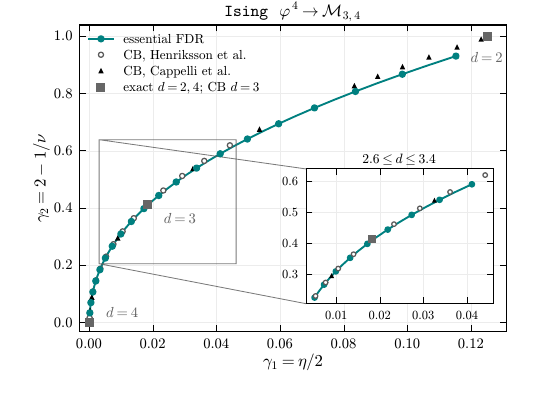}\hfill\includegraphics[width=0.49\textwidth,trim=0 16 0 0,clip]{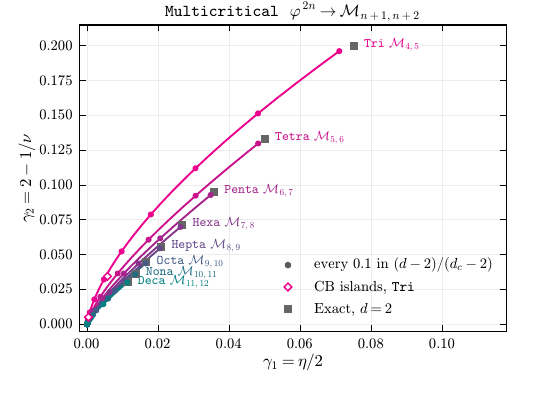}
    \caption{The multicritical classes in the plane of the anomalous dimensions $\gamma_1=\eta/2$, $\gamma_2=2-1/\nu$, parametric
    in $d$, as in the bootstrap study of Ref.~\cite{El-Showk:2013nia}. Left: {\tt Ising} from $d=4$ to $d=2$, with the bootstrap
    (CB) in fractional $d$ of Ref.~\cite{Henriksson:2022gpa} ($2.6\le d\le3.95$, open circles) and of
    Refs.~\cite{Cappelli:2018vir,Bonanno:2022ljl} ($2<d<4$, triangles); squares: the exact values at $d=4$ and $d=2$ and the best CB value at $d=3$ ($\Delta_\sigma=0.518148806$,
    $\Delta_\epsilon=1.41262528$); dots every $0.1$ in $d$; inset: $2.6\le d\le3.4$. Right: $\varphi^{2n}$, $n=3,\dots,10$, from $d_c$
    to $d=2$; dots every $0.1$ in $x=(d-2)/(d_c-2)$; diamonds: the {\tt Tricritical} bootstrap islands in $d=2.75$ and
    $2.5$ \cite{Henriksson:2025tri}; squares: exact CMM values at $d=2$.}
    \label{G1G2}
\end{figure*}

The field expansion is adequate in $d=3$, where the Wilson--Fisher fixed point is the only non-trivial scaling solution.
Below $d=3$ the multicritical fixed points appear one after the other, and following them requires a functional solution. Being one-dimensional, Eqs.~\eqref{betav}--\eqref{betaf} can be studied directly at the functional level, in every $d$. Defining
$u(\varphi) = v'(\varphi)$ and
$Q\equiv(f'+\tfrac{d+2}{2})u+(f-\tfrac{d-2}{2}\varphi)u'$, the fixed-point equation for $u$ and the derivative of the constraint
\eqref{betaf} read
\begin{eqnarray}
0&=& u''+Q\,e^{u'}\,,\label{equ}\\
0&=&f''+\tfrac{1}{12}\,Q^2u''e^{u'}+\tfrac{1}{6}\,Q\,Q'e^{u'}\,,\nonumber
\end{eqnarray}
with $Q'=uf''+2(1+f')u'+(f-\tfrac{d-2}{2}\varphi)u''$. The initial conditions for the implicit system \eqref{equ} are
$u(0)=0$, $u'(0)=\sigma$ and $f(0)=f'(0)=0$. A generic solution breaks down at a finite field $\varphi_*(\sigma)$; global scaling
solutions are the values of $\sigma$ at which $\varphi_*\to\infty$, the spikes of the plot $\varphi_*(\sigma)$ \cite{Morris:1994ki,Codello:2012sc,Hellwig:2015woa}. The
search for fixed points is thus a one-parameter shooting problem, exactly as in the LPA. In the limit $d\to2$ we expect a countable infinity of spikes, one for each minimal model.

Figure~\ref{EssentialSpike} shows the spike plots in $d=3$, $2.45$ and $2$. In $d=3$ there are two spikes: {\tt Ising} at
$\sigma_{\tt Ising} = -0.26580$, the same value found with the field expansion, and the Gaussian at $\sigma=0$. In $d=2.45$, between
the upper critical dimensions of $\varphi^{10}$ and $\varphi^{8}$, {\tt Tri} and {\tt Tetra} have joined, on opposite sides of
$\sigma=0$. The fixed-point potential of the $\varphi^{2n}$ class, obtained by shooting at $\sigma_*$, has $n$ minima; the field
redefinition $f$ is small and odd.

In $d=2$ we find exactly what conformal field theory predicts: the spike plot contains all the multicritical fixed points, one for
each minimal model. To our knowledge, this is the first one-dimensional spike plot computed directly in two dimensions. The spikes alternate
between the two sides of $\sigma=0$, with {\tt Ising}, {\tt Tetra}, {\tt Hexa}, $\dots$ at $\sigma<0$ and {\tt Tri}, {\tt Penta},
$\dots$ at $\sigma>0$, and accumulate on the Gaussian point. Each spike ends a step of a staircase, the CMM stairway of
Fig.~\ref{EssentialStairway}. The potential of $\mathcal{M}_{m,m+1}$ has $m-1$ minima, evenly spaced by $\Delta\varphi\simeq4.4$ (approaching $4.44$ as $m$
grows), and along both branches $|\sigma_*|$ decreases approximately as $1/n$.

Stacking the spike plots in $d$ gives the {\color{teal}{\it Cordillera of} CMM} (Fig.~\ref{CordilleraShadow}, inset). Each multicritical class is a
ridge that leaves the Gaussian ridge $\sigma=0$ at its $d_c=\frac{2n}{n-1}$, where $\varphi^{2n}$ becomes marginal, and continues
smoothly down to $d=2$. Its shadow on the $(\sigma,d)$ plane (Fig.~\ref{CordilleraShadow}, main panel) coincides with the lines of
fixed points $\sigma_*(d)=-\ln(d\,v_*(0))$ obtained by continuation in $d$, and each line ends, at $d=2$, on its minimal model
$\mathcal{M}_{n+1,n+2}$. The whole family of multicritical universality classes, from the upper critical dimensions to the
two-dimensional CFTs, is thus contained in the solutions of Eqs.~\eqref{betav}--\eqref{betaf}.

\paragraph*{\color{teal}\mbox{Spectrum Continuity.---}}
\begin{table}[t]
\footnotesize
\setlength{\tabcolsep}{2.6pt}
\renewcommand{\arraystretch}{1.15}
\begin{tabular}{c|r|ll|ll|ll}
CFT & $\sigma_*$ & $\eta$ & $\eta_{{\rm ex}}$ & $\nu$ & $\nu_{{\rm ex}}$ & $\omega$ & $\omega_{{\rm ex}}$\tabularnewline
\hline
$\mathcal{M}_{3,4}$ & $-0.77171$ & $0.2302$ & $0.25$ & $0.9358$ & $1$ & $1.792$ & $2$\tabularnewline
$\mathcal{M}_{4,5}$ & $1.11331$ & $0.1419$ & $0.15$ & $0.5544$ & $0.5556$ & $1.160$ & $1$\tabularnewline
$\mathcal{M}_{5,6}$ & $-0.53202$ & $0.0963$ & $0.1$ & $0.5347$ & $0.5357$ & $0.877$ & $0.8$\tabularnewline
$\mathcal{M}_{6,7}$ & $0.66854$ & $0.0695$ & $0.07143$ & $0.5244$ & $0.525$ & $0.710$ & $0.6667$\tabularnewline
$\mathcal{M}_{7,8}$ & $-0.40592$ & $0.0525$ & $0.05357$ & $0.5181$ & $0.5185$ & $0.599$ & $0.5714$\tabularnewline
$\mathcal{M}_{8,9}$ & $0.48012$ & $0.0411$ & $0.04167$ & $0.5140$ & $0.5143$ & $0.519$ & $0.5$\tabularnewline
$\mathcal{M}_{9,10}$ & $-0.32835$ & $0.0330$ & $0.03333$ & $0.5112$ & $0.5114$ & $0.458$ & $0.4444$\tabularnewline
$\mathcal{M}_{10,11}$ & $0.37510$ & $0.0270$ & $0.02727$ & $0.5091$ & $0.5093$ & $0.410$ & $0.4$\tabularnewline
$\mathcal{M}_{11,12}$ & $-0.27577$ & $0.0226$ & $0.02273$ & $0.5076$ & $0.5077$ & $0.371$ & $0.3636$\tabularnewline
\end{tabular}
\caption{Critical exponents of the CMM $\mathcal{M}_{n+1,n+2}$ ({\tt Ising}, {\tt Tri}, \dots, {\tt Deca}) from
essential FDR-DE2 eigen-perturbations directly in $d=2$, with the exact values $\eta_{\rm ex}=3/[m(m+1)]$,
and, for $m\ge4$, $\nu_{\rm ex}=m(m+1)/[2(m-1)(m+2)]$, $\omega_{\rm ex}=4/m$ ($1$ and $2$ for {\tt Ising}). Numerical uncertainty: $\sigma_*$,
$\eta$ and $\nu$ below $2\times10^{-5}$; $\omega$ about $\pm10^{-3}$ for $m\le7$ and $\pm10^{-4}$ above.}
\label{tab:d2}
\end{table}

Around a scaling solution, the eigen-perturbations $\delta v\propto e^{\theta t}$ and $\delta f$ solve a linear third-order system for
$(\delta u,\delta u',\delta f)$, with even ($\delta v$ even, $\delta f$ odd) and odd ($\delta v$ odd, $\delta f$ even) sectors. The
eigenvalues follow from matching the solutions integrated from the origin and from the singular point of the system (End Matter), and the scaling dimensions are $\Delta=d+\theta$. We extract
$\nu=-1/\theta_1$ from the most negative even eigenvalue, $\omega$ as the smallest positive even eigenvalue, and
$\eta=d+2+2\theta_h$ from the most negative odd eigenvalue $\theta_h$, excluding the redundant shift mode $\theta=-\frac{d-2}{2}$
(the descendant $\partial^2\varphi$, with $\Delta=\frac{d+2}{2}$ exactly).

Figure~\ref{CMMspectral} shows the spin-zero spectra of the first nine classes between $d=2$ and $d_c$. All eigenvalues stay
real: in the essential scheme there are no redundant operators to collide with, and spectrum continuity is restored. Each level
interpolates between the Gaussian operator $\varphi^k$ at $d_c$, with $\Delta_k=\frac{k}{n-1}$, and a Kac primary of the minimal model at $d=2$:
$\varphi^k\to\Phi_{k+1,k+1}$ for $k\le m-2$ and $\varphi^{k}\to\Phi_{k-m+3,k-m+2}$ for $m-1\le k\le 2m-4$, with
$\Delta_{r,s}=2h_{r,s}=\frac{((m+1)r-ms)^2-1}{2m(m+1)}$. The RG thus reconstructs, level by level and without any input from the
Virasoro algebra, the identification of the Landau--Ginzburg operators with the Kac table \cite{Zamolodchikov:1986db}.

The critical exponents in $d=2$ are collected in Table~\ref{tab:d2}. The relative error of $\eta$ falls from $-7.9\%$ for
{\tt Ising} to $-0.7\%$ for $\mathcal{M}_{11,12}$, and that of $\nu$ from $-0.2\%$ for {\tt Tri} to $-0.02\%$; the errors of all the
primaries stay below $8\%$ and decrease with $m$ (End Matter, Figs.~\ref{etanuvsm}--\ref{relerr}). This is not obvious: as $m\to\infty$, $d_c\to2$ and
the fixed points approach the Gaussian one, a limit that is perturbative in Zamolodchikov's two-dimensional formulation
\cite{Zamolodchikov:1987ti} but strongly non-perturbative in the Landau--Ginzburg field basis used here. Indeed, in perturbation theory the $\varphi^{2n}$ fixed point first appears at $n-1$ loops and
the $\varepsilon$-expansion does not reach $d=2$ \cite{ODwyer:2007brp,Itzykson:1989sx}. The FDR flow, which sums all critical
dimensions, captures every multicritical class already at one loop (End Matter).

Away from $d=2$ the decisive test is the comparison with the bootstrap spectra in fractional dimensions
\cite{Henriksson:2022gpa,Henriksson:2025tri} (End Matter, Figs.~\ref{SpikeDE2}--\ref{TriComp}). For {\tt Ising}, the levels
$\varphi$, $\varphi^2$, $\varphi^4$, $\varphi^5$ and $\varphi^6$ follow the spin-zero bootstrap spectrum of
Ref.~\cite{Henriksson:2022gpa} within $1$--$2\%$ over the whole range $2.6\le d\le3.95$; the bootstrap levels that do not follow a
$\varphi^k$ curve are operators with derivatives, absent at DE2. The bootstrap of Refs.~\cite{Cappelli:2018vir,Bonanno:2022ljl}
extends the comparison down to $d=2$, where $\epsilon''$ and our $\varphi^6$ level reach $\Delta=8$ together. For {\tt Tricritical}, the bootstrap islands of
Ref.~\cite{Henriksson:2025tri} in $d=2.75$ and $d=2.5$ lie on our curves. Figure~\ref{G1G2} summarizes the result: each class traces a smooth curve in the
plane of the anomalous dimensions $\gamma_1=\eta/2$, $\gamma_2=2-1/\nu$, from the Gaussian point at $d_c$ to its minimal model at $d=2$.

\paragraph*{\color{teal}\mbox{Outlook.---}}

Two RG flow equations, Eqs.~\eqref{betav}--\eqref{betaf}, derived at one loop and with no input from conformal symmetry, reproduce
the structure of all the CMM and, with high precision, their low-lying spin-zero spectrum, following each of them continuously from
its upper critical dimension down to two dimensions. Two ingredients make this possible: FDR,
which keeps the flow simple without losing precision and gives access to $d=2$, and the essential scheme, which removes the redundant operators and keeps the
spectrum real. Our key results are three. First, we establish the continuity of the fixed points and of their low-lying spin-zero
spectrum for a general multicritical model, beyond the {\tt Ising} and {\tt Tricritical} cases known so far, thus completing the
construction of the CMM over their whole range of existence $2<d<d_c$. Second, we show that FDR is a field-theoretic RG that
overcomes the limitations of the $\varepsilon$-expansion and reaches two dimensions without difficulty: two-dimensional CFTs can be
seen accurately with a one-loop approach. Third, we demonstrate that essentiality is the key to spectrum continuity, in FDR and, we expect, in any functional RG that keeps redundant operators.

The limits of the present description are equally clear. Operators with derivatives, and hence higher spins, lie beyond the second
order of the derivative expansion: they appear in the bootstrap spectrum as additional levels, which can repel those of the
potential \cite{Henriksson:2022gpa}. The fourth order is the natural next step. Furthermore, the extension of FDR beyond one loop is
another direction to pursue, which may also clarify why the $O(\varepsilon^2)$ terms are exact only for $d_c=4$.

That such a simple RG equation captures the CMM makes it a tool for exploration. Natural next steps are the non-unitary conformal minimal models,
whose $d\to2$ limit has been matched with CFTs only for {\tt Lee-Yang}
\cite{Beretta:2026ly,Kockler:2026yle,Benedetti:2026lyf}, and the D-series of minimal models \cite{Cappelli:2009xj}, such as the $3$-state Potts model and
its generalizations, whose Landau--Ginzburg description is possible but more problematic, with larger quantum corrections. Further
targets are the global phase diagrams of the $O(N)$ models in the $(d,N)$
plane \cite{Codello:2012ON,Codello:2014ON,Jakubczyk:2014KT,Chlebicki:2021an,Chlebicki:2023lt,Henriksson:2022rev,DeCesare:2025nlsm,Beretta:2026on}
and of the Potts models in the $(d,Q)$ plane \cite{BenAliZinati:2017pt,Chester:2022pt,SanchezVillalobos:2023pt,Wiese:2023pt}, and
more general multi-scalar theories \cite{Osborn:2017ucf,BenAliZinati:2019pft}. More broadly, it opens the way to a classification and construction of
interacting field theories in all dimensions on RG grounds alone, complementary to the CB.

\paragraph*{\color{teal}\mbox{Acknowledgments.---}}

I thank R.~Ben~Al\`i~Zinati and K.~Falls for introducing me to the essential scheme and for many fruitful discussions, and the
organizers and participants of the winter school {\it Modern Applications of Field Theory} (IESC Carg\`ese, Corsica, 9--13 November
2026), where these results were first presented, in particular A.~Cappelli and A.~Trombettoni. AI tools have been used for typesetting, proofreading and plotting, as well as for optimizing numerical scripts.

\section*{End Matter}

\paragraph*{\color{teal}\mbox{Recovering the $\varepsilon$-Expansion.---}}

\begin{figure}[t]
    \centering
    \includegraphics[width=\columnwidth]{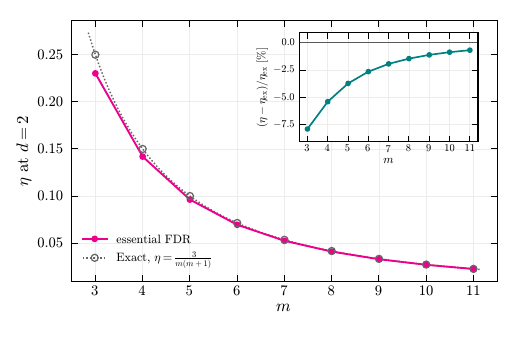}\\
    \includegraphics[width=\columnwidth,trim=0 13 0 0,clip]{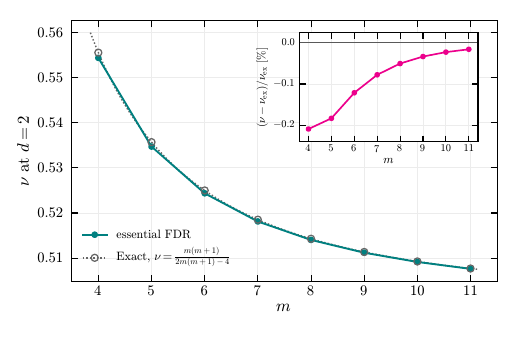}
    \caption{$\eta$ (top) and $\nu$ (bottom) at $d=2$ vs $m$, for $\varphi^{2n}\to\mathcal{M}_{m,m+1}$ with $m=n+1$. Open circles:
    exact values; dotted: the exact $\eta=\frac{3}{m(m+1)}$ and $\nu=1/(2-\Delta_\epsilon)$, $\Delta_\epsilon=\frac{4}{m(m+1)}$, continued
    in $m$; $\nu$ is shown for $m\ge4$ ({\tt Ising}, $\nu=0.9358$ against $1$, is off scale). Insets: relative error
    $(X-X_{\rm ex})/X_{\rm ex}$ in per cent, which tends to zero as $m\to\infty$.}
    \label{etanuvsm}
\end{figure}
\begin{figure*}[t]
    \centering
    \includegraphics[width=\textwidth,trim=0 14 0 0,clip]{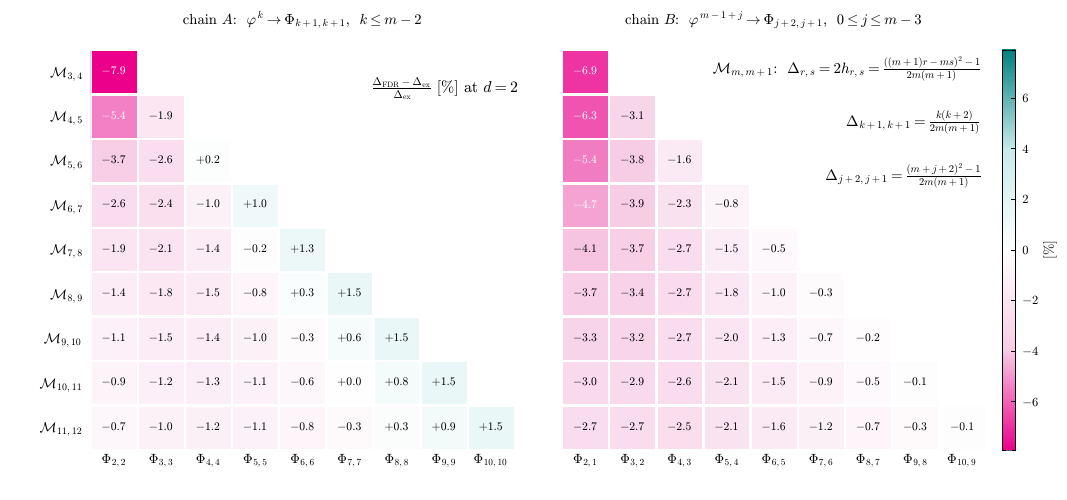}\\[1.5mm]
    \caption{Relative error $(\Delta_{\rm FDR}-\Delta_{\rm ex})/\Delta_{\rm ex}$ in per cent of every primary $\varphi^k$,
    $k\le2n-2$, at $d=2$. Rows: $\mathcal{M}_{m,m+1}$, $m=n+1$; columns: the Kac operator, ordered along the two chains $A$,
    $\varphi^k\to\Phi_{k+1,k+1}$ ($k\le m-2$), and $B$, $\varphi^{m-1+j}\to\Phi_{j+2,j+1}$ ($0\le j\le m-3$); the exact dimensions are
    given in the upper right.}
    \label{relerr}
\end{figure*}
\begin{figure*}[t]
    \centering
    \includegraphics[width=0.46\textwidth,trim=0 13 0 0,clip]{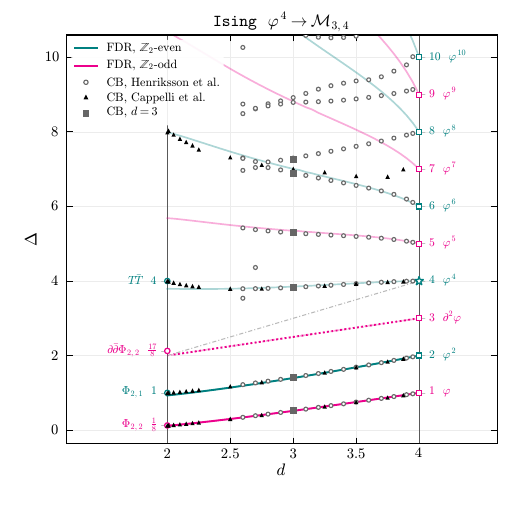}\qquad\includegraphics[width=0.46\textwidth,trim=0 13 0 0,clip]{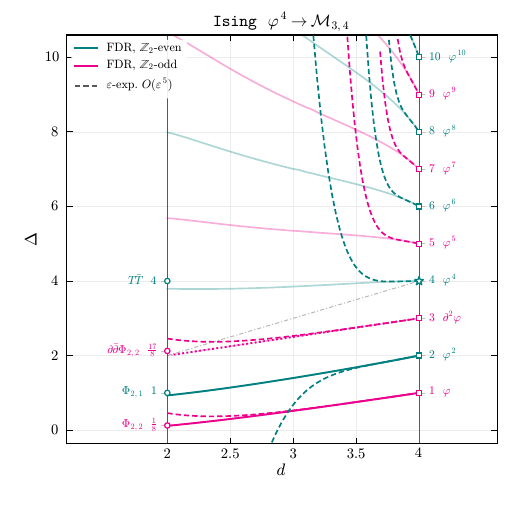}
    \caption{Spectrum continuity of the spin-zero operators of the {\tt Ising} class between $2\leq d \leq 4$: essential FDR
    spectrum of Fig.~\ref{CMMspectral} against reference data. Left: the spin-zero bootstrap spectrum in fractional $d$
    \cite{Henriksson:2022gpa} ($2.6\le d\le3.95$; open circles), the bootstrap dimensions of
    $\sigma,\epsilon,\epsilon',\epsilon''$ for $2<d<4$ \cite{Cappelli:2018vir,Bonanno:2022ljl} (triangles; Table~1 of
    Ref.~\cite{Bonanno:2022ljl}), and the best $d=3$ bootstrap values of $\sigma,\epsilon,\epsilon',\sigma',\epsilon'',\epsilon'''$
    (squares). The bootstrap levels not following a $\varphi^k$ curve are operators with derivatives, absent at DE2. Right: five-loop
    $\varepsilon$-expansion of $\Delta_{\varphi^k}$, $k=1,2,4,\dots,10$ \cite{Henriksson:2025hwi,Henriksson:2025mls}, plain truncated
    series without resummation (dashed); the descendant is $\Delta_\varphi+2$. Beyond $\varphi^2$ the series leaves the plot within
    $\varepsilon\lesssim1$.}
    \label{SpikeDE2}
\end{figure*}
\begin{figure*}[t]
    \centering
    \includegraphics[width=0.46\textwidth,trim=0 13 0 0,clip]{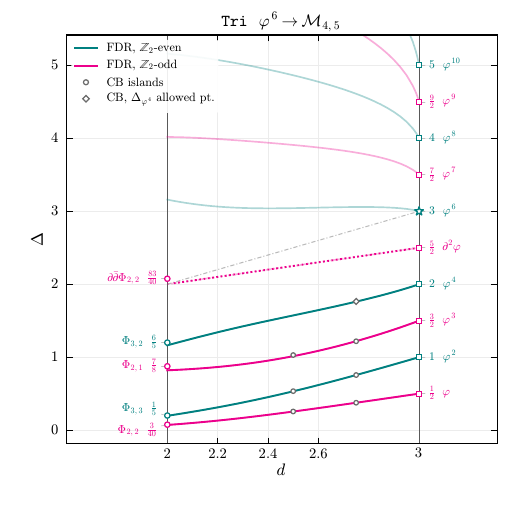}\qquad\includegraphics[width=0.46\textwidth,trim=0 13 0 0,clip]{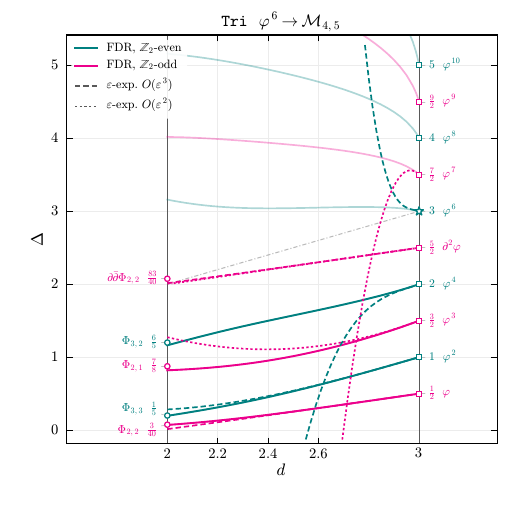}
    \caption{Spectrum continuity of the spin-zero operators of the {\tt Tricritical} class between $2\leq d \leq 3$. Left: the
    three-correlator bootstrap islands in $d=2.75$ and $d=2.5$ for $\Delta_\varphi$, $\Delta_{\varphi^2}$, $\Delta_{\varphi^3}$ (centre of
    each island, whose extent is smaller than the marker), and $\Delta_{\varphi^4}=1.7637$ at an allowed point in $d=2.75$
    \cite{Henriksson:2025tri}; no islands were found in $d=2.25$ and $d=2$. Right: $\varepsilon$-expansion in $d=3-\varepsilon$,
    $\Delta_{\varphi^k}=k(1-\varepsilon)/2+\gamma_k$, plain truncated series: $k=1,2,4,6$ to $O(\varepsilon^3)$ (dashed) from the corrected
    six-loop results \cite{Adzhemyan:2026six,Jack:2026phi6} (Appendix B of Ref.~\cite{Henriksson:2025tri}), and $k=3,7$ to
    $O(\varepsilon^2)$ (dotted) \cite{Henriksson:2025tri}; $\varphi^5=\partial^2\varphi$ is $\Delta_\varphi+2$.}
    \label{TriComp}
\end{figure*}
\begin{figure*}[t]
\centering
\includegraphics[width=0.46\textwidth]{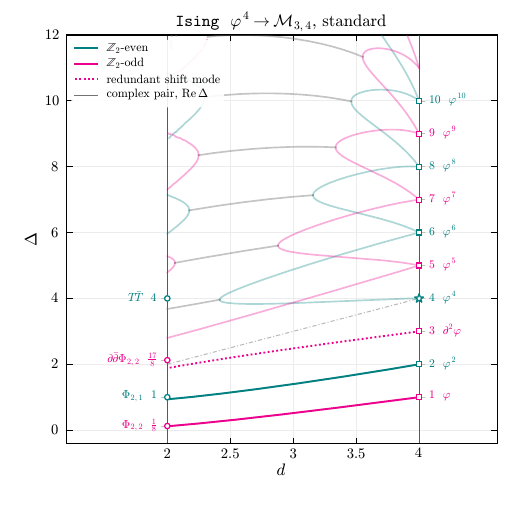}\qquad\includegraphics[width=0.46\textwidth]{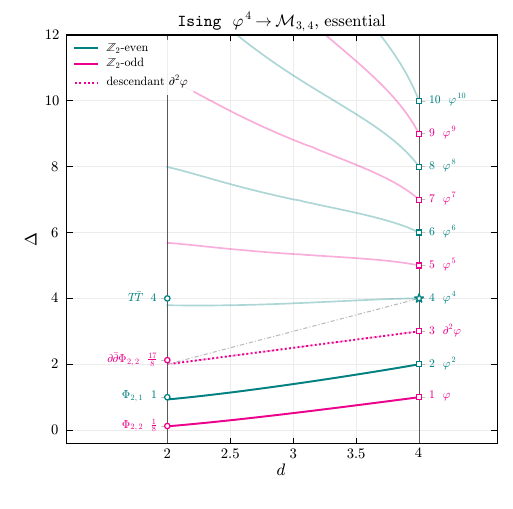}\\[1mm]
\includegraphics[width=0.46\textwidth,trim=0 13 0 0,clip]{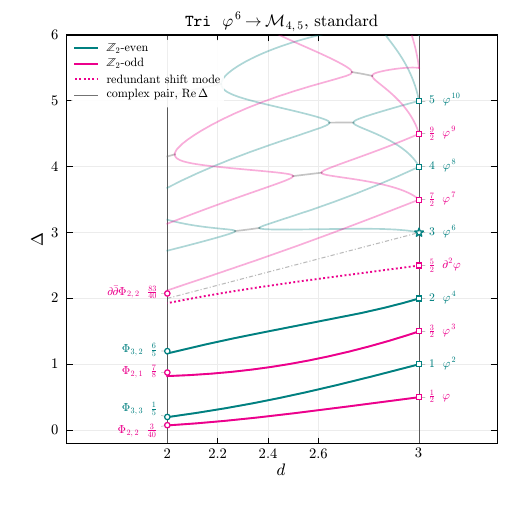}\qquad\includegraphics[width=0.46\textwidth,trim=0 13 0 0,clip]{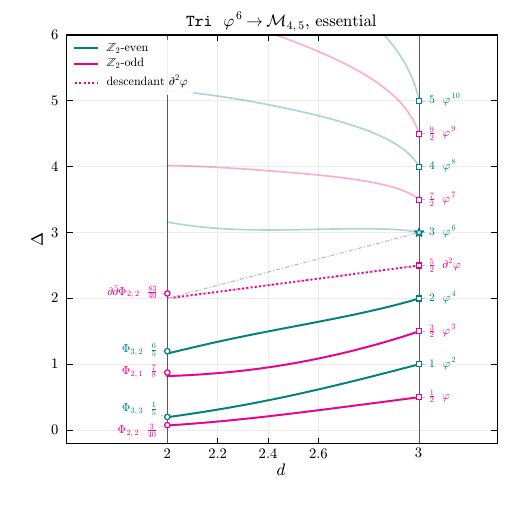}
\caption{Spin-zero scaling dimensions of {\tt Ising} (top) and {\tt Tricritical} (bottom) between $d=2$ and $d_c$, in
the standard (left) and essential (right) FDR-DE2 schemes; $\mathbb{Z}_2$-even in teal, odd in magenta, light where irrelevant. In the standard
scheme levels collide pairwise and become complex (grey: ${\rm Re}\,\Delta$); each collision is joined at its branch point,
located by extrapolating $(\Delta_+-\Delta_-)^2$ linearly to zero. Dotted: the redundant shift mode (standard) and the descendant
$\partial^2\varphi$ (essential). Gaussian operators $\varphi^k$ at $d_c$ (squares, right) and exact two-dimensional values (circles,
left).}
\label{SpectrumDE2}
\end{figure*}

Starting from the field expansion~\eqref{fexpbeta}--\eqref{fexpcond}, with the $\gamma$'s eliminated as described in the main text,
we can solve all couplings in terms of the first two, $\chi \equiv \lambda_1$ and $\sigma \equiv \lambda_2$.
We use $\beta_1=0$ to solve for $\lambda_3(\chi,\sigma)$, insert it into $\beta_2=0$ to solve for $\lambda_4(\chi,\sigma)$, and so on up
to $\beta_{n-2} = 0$ for $\lambda_n(\chi,\sigma)$. Around $d=d_c-\varepsilon$ we expect $\chi= \beta \varepsilon$ and $\sigma= \alpha
\varepsilon$, and we then expand $\beta_{n-1} = 0$ and $\beta_{n} = 0$ in $\varepsilon$. For $d_c=4$ we find $\chi=0$ and $\alpha =-
\tfrac{1}{6}$, which leads to $\lambda_2 = -\tfrac{\varepsilon}{6}$, $\lambda_4 = \tfrac{\varepsilon}{3}$ and $\lambda_i =
O(\varepsilon^2)$ for $i=1,3$ and $i>4$. The stability matrix $\partial\beta_i/\partial\lambda_j|_*$ then gives
\begin{eqnarray}
\Delta_1 &=& 1-\tfrac{\varepsilon}{2}+\tfrac{\varepsilon^2}{108}\,,\qquad
\Delta_2 = 2-\tfrac{2\varepsilon}{3}\,,\nonumber\\
\Delta_3 &=& 3-\tfrac{\varepsilon}{2}-\tfrac{\varepsilon^2}{108}\,,\qquad
\Delta_4 = 4 +O(\varepsilon^2)\,,
\end{eqnarray}
up to $O(\varepsilon^3)$ for $\Delta_{1,3}$ and $O(\varepsilon^2)$ for $\Delta_2$, or, in compact form,
\begin{equation}\label{epsDelta}
\Delta_n = n\,\tfrac{d-2}{2} + \tfrac{n(n-1)}{6}\, \varepsilon +O(\varepsilon^2)\,,
\end{equation}
which agrees at $O(\varepsilon)$ with the known results \cite{ODwyer:2007brp,Codello:2017hhh} for all $n$. Furthermore we capture the
correct $O(\varepsilon^2)$ terms of $\Delta_1$ and $\Delta_3$. In particular $\eta=2\Delta_1-d+2=\tfrac{\varepsilon ^2}{54}+O(\varepsilon^3)$
is the correct two-loop anomalous dimension, and the relation $\Delta_1+\Delta_3=d$ holds.

The same result follows at the functional level \cite{ODwyer:2007brp,Benedetti:2026lyf}, in every critical dimension. Below $d_c=\frac{2n}{n-1}$,
the fixed point is
\begin{eqnarray}
v_*&=&\tfrac{1}{d_c}+\varepsilon\Big\{A_n{\rm He}_{2n}\big(\tfrac{\varphi}{\sqrt{n-1}}\big)+\tfrac{1}{d_c^{2}}\Big\}+O(\varepsilon^2)\,,\nonumber\\
A_n&=&\tfrac{2(n-1)^3(n-2)!\,(n!)^2}{[(2n)!]^2}\,,
\end{eqnarray}
and $f_*=O(\varepsilon^2)$, with $f_2'=-\frac{1}{12}(v_1''')^2$. For $n=2$, $\sigma_*=-\frac{\varepsilon}{6}$; for $n=3$,
$\sigma_*=\frac{\varepsilon}{20}$. We checked that $\sigma_*(d)/\varepsilon$ from the numerical continuation tends to these values as
$\varepsilon\to0$ for $n=2,\dots,5$. In $d_c=3,\tfrac83,\dots$ the $\varepsilon$-expansion is thus recovered at leading order, but the $O(\varepsilon^2)$
coefficients are no longer exact: for example, $\eta=\frac{\varepsilon^2}{450}$ versus the exact $\frac{\varepsilon^2}{500}$ for
$d_c=3$, the same value found in the standard FDR-DE2 scheme \cite{FDR}.

\paragraph*{\color{teal}\mbox{Eigen-Perturbations.---}}

Writing $v=v_0+\epsilon\,\delta v\,e^{\theta t}$, $f=f_0+\epsilon\,\delta f\,e^{\theta t}$ around a scaling solution $v_0$, $f_0$,
with $u_0=v_0'$, $\delta u=\delta v'$ and $E_0=e^{-u_0'}$, the linearized flow and constraint read
\begin{eqnarray}
\theta\,\delta u&=&E_0\Big(\tfrac{u_0u_0''}{6}-1\Big)\delta u''+\Big[-\tfrac{d+2}{2}+\tfrac{E_0}{12}(u_0'')^2\Big]\delta u\nonumber\\
&&+\Big[\tfrac{d-2}{2}\varphi-f_0+E_0u_0''-\tfrac{u_0E_0}{12}(u_0'')^2\Big]\delta u'-u_0'\,\delta f\,,\nonumber\\
\delta f'&=&-\tfrac{E_0}{12}\big[2u_0''\,\delta u''-(u_0'')^2\,\delta u'\big]\,.
\end{eqnarray}
In the even sector $\delta u(0)=\delta f(0)=0$; in the odd sector $\delta u'(0)=\delta f(0)=0$. The coefficient of $\delta u''$ vanishes at the point $\varphi_c$ where $u_0u_0''=6$,
a regular singular point: admissible eigenfunctions span the two-dimensional regular subspace there, which replaces the condition at
infinity (beyond $\varphi_c$ the fast mode decays). The eigenvalues are the zeros of the Evans function \cite{Evans:1975,Alexander:1990} built from the solution
integrated outward from the origin and the two regular solutions integrated inward from $\varphi_c$, matched at
$\varphi_m<\varphi_c$; they are independent of the numerical parameters to $\sim10^{-9}$, and the exact redundant shift mode
$\delta v=v_0'$, $\delta f=f_0'$ ($\theta=-\frac{d-2}{2}$) is reproduced to the same accuracy.

\paragraph*{\color{teal}Comparison with Exact, CB and $\varepsilon$-Expansion.---}

In $d=2$ the critical exponents of Table~\ref{tab:d2} are compared with the exact values in Fig.~\ref{etanuvsm}, and the
errors of all the primaries $\varphi^k$, $k\le2n-2$, are shown in Fig.~\ref{relerr}. The agreement is already good for {\tt Ising},
with all errors below $8\%$, and it improves steadily with the order of multicriticality: the errors decrease with $m$, down to $0.7\%$ for $\eta$ and $0.02\%$ for $\nu$ in $\mathcal{M}_{11,12}$. A possible
explanation is the following. As $n\to\infty$, $d_c\to2$ and the fixed points approach the Gaussian one. In standard perturbation
theory, however, the $\varphi^{2n}$ fixed point below $d_c$ first appears at $n-1$ loops, so that higher multicriticality requires
higher loop orders, and the $\varepsilon$-expansion does not extend down to $d=2$ even though the fixed point becomes increasingly
Gaussian \cite{ODwyer:2007brp,Itzykson:1989sx}. In FDR, instead, the sum over all critical dimensions brings the contributions of
every $\varphi^{2n}$ into the one-loop flow: all multicritical fixed points are captured at once, and their description improves as
they approach the Gaussian one.

Above $d=2$ the spectrum can be compared with the CB in fractional dimensions \cite{El-Showk:2013nia} and with the $\varepsilon$-expansion
(Figs.~\ref{SpikeDE2} and \ref{TriComp}). For {\tt Ising} the levels $\varphi$, $\varphi^2$, $\varphi^4$, $\varphi^5$ and $\varphi^6$
follow the spin-zero CB spectrum of Ref.~\cite{Henriksson:2022gpa} in $2.6\le d\le3.95$ within $1$--$2\%$. The CB levels that do not
follow a $\varphi^k$ curve are operators with derivatives, absent at DE2; one of them, $\partial^4\varphi^4$, repels the $\varphi^6$
level below $d\simeq2.8$ \cite{Henriksson:2022gpa}. The bootstrap of Refs.~\cite{Cappelli:2018vir,Bonanno:2022ljl} covers the
whole range $2<d<4$: $\sigma$, $\epsilon$ and $\epsilon'$ agree with our levels within $1\%$ for $d\ge3$, the deviations growing
smoothly to their two-dimensional values ($5$--$8\%$), and for $d\lesssim3.3$ $\epsilon''$ follows our $\varphi^6$ level within $3\%$.
For {\tt Tricritical} the bootstrap islands of Ref.~\cite{Henriksson:2025tri} in
$d=2.75$ and $d=2.5$ lie on our curves. The unresummed $\varepsilon$-expansion, to $O(\varepsilon^5)$ for {\tt Ising}
\cite{Henriksson:2025hwi,Henriksson:2025mls} and to $O(\varepsilon^3)$ for {\tt Tricritical} \cite{Henriksson:2025tri,Adzhemyan:2026six,Jack:2026phi6}, follows the essential FDR curves near
$d_c$ and departs from them for the higher operators well before $d=2$.

\paragraph*{\color{teal}\mbox{Failure of the Standard Scheme.---}}

Here we show that the standard scheme breaks spectrum continuity also in FDR. This scheme keeps the
wave-function renormalization $Z(\phi)$, with dimensionless $z(\varphi)$ normalized by $z(0)=1$, and the FDR-DE2 flow reads
\cite{FDR}
\begin{eqnarray}
\partial_t v&=&-d\,v+\tfrac{d-2+\eta}{2}\varphi v'+e^{-v''/z}\,,\label{stdflow}\\
\partial_t z&=&\eta z+\tfrac{d-2+\eta}{2}\varphi z'+e^{-v''/z}\Big\{-\tfrac{(v''')^2}{6z^2}-\tfrac{z''}{z}\nonumber\\
&&+\Big[\tfrac{1}{z^2}+\tfrac{v''}{3z^3}\Big]v'''z'
+\Big[\tfrac{3}{2z^2}-\tfrac{v''}{z^3}-\tfrac{(v'')^2}{6z^4}\Big]z'^2\Big\}\,,\nonumber
\end{eqnarray}
where $\eta=-\{\cdots\}e^{-v''/z}\,|_{\varphi=0}$ preserves $z(0)=1$. Trading $z$ for the field redefinition $f$ gives back the
essential equations \eqref{betav}--\eqref{betaf}. The scaling solutions of Eq.~\eqref{stdflow}, instead, are the peaks of a
landscape over the $(\sigma,\eta)$ plane \cite{FDR2d}, and their spectrum, acting on $(\delta v,\delta z)$, contains redundant
directions besides the essential levels: the shift mode, with $\Delta=\frac{d+2-\eta}{2}$, and further levels absent from the
essential spectrum. Following the fixed points in $d$, these levels cross the essential ones as $d$ decreases, and since the
stability matrix is not symmetric, levels of the same parity that meet collide and become complex conjugate pairs
(Fig.~\ref{SpectrumDE2}). For {\tt Ising} the $\varphi^4$ level collides at $d\simeq2.42$, so that $\omega$ is complex for all
$d<2.42$, with ${\rm Im}\,\Delta=0.24$ at $d=2$; higher pairs are complex over finite windows. For {\tt Tricritical} the marginal
$\varphi^6$ level is complex for $2.27<d<2.37$. In the essential scheme, instead, the field redefinition removes the redundant
directions and all levels stay real.

\end{document}